\let\ifarxiv=\iftrue     
\pdfoutput=1
{\def\usepackage{ws-procs9x6}}

\ifarxiv

\documentclass[12pt,a4paper]{article}
\usepackage[a4paper,text={450pt,650pt},centering]{geometry}

\fi

\ifarxiv\else

\documentclass[11pt,a4paper]{article}
\usepackage{mathptmx}
\usepackage[a4paper,text={130mm,198mm}]{geometry}

\fi

\ifnum\pdfoutput=1\else
\PassOptionsToPackage{hypertex}{hyperref}
\PassOptionsToPackage{draft}{graphicx}
\usepackage{showkeys}
\fi

\usepackage{amsmath,amssymb}
\usepackage[bookmarks=true,hyperfigures=true]{hyperref}
\usepackage{graphicx}
\usepackage[nosort]{cite}
\ifarxiv\usepackage[bulletsep]{collref}\fi

\let\oldbfseries=\bfseries
\let\oldmdseries=\mdseries
\let\oldnormalfont=\normalfont
\renewcommand{\bfseries}{\oldbfseries\boldmath}
\renewcommand{\mdseries}{\oldmdseries\unboldmath}
\renewcommand{\normalfont}{\oldnormalfont\unboldmath}

\allowdisplaybreaks[3]

\numberwithin{equation}{section}

\usepackage[font=small,labelfont=bf,width=0.85\textwidth]{caption}

\providecommand{\hypersetup}[1]{}
\providecommand{\texorpdfstring}[2]{#1}

\hypersetup{plainpages=false}
\hypersetup{pdfpagemode=UseNone}
\hypersetup{bookmarksnumbered=true}
\hypersetup{pdfstartview=FitH}
\hypersetup{colorlinks=false}
\hypersetup{citebordercolor={.5 1 .5}}
\hypersetup{urlbordercolor={.5 1 1}}
\hypersetup{linkbordercolor={1 .7 .7}}

\DeclareMathSymbol{\Gamma}{\mathalpha}{letters}{"00}
\DeclareMathSymbol{\Delta}{\mathalpha}{letters}{"01}
\DeclareMathSymbol{\Theta}{\mathalpha}{letters}{"02}
\DeclareMathSymbol{\Lambda}{\mathalpha}{letters}{"03}
\DeclareMathSymbol{\Xi}{\mathalpha}{letters}{"04}
\DeclareMathSymbol{\Pi}{\mathalpha}{letters}{"05}
\DeclareMathSymbol{\Sigma}{\mathalpha}{letters}{"06}
\DeclareMathSymbol{\Upsilon}{\mathalpha}{letters}{"07}
\DeclareMathSymbol{\Phi}{\mathalpha}{letters}{"08}
\DeclareMathSymbol{\Psi}{\mathalpha}{letters}{"09}
\DeclareMathSymbol{\Omega}{\mathalpha}{letters}{"0A}

\newcommand{\gen}[1]{\mathrm{#1}}
\newcommand{\superN}{\mathcal{N}}
\newcommand{\Op}[1]{\mathcal{#1}}
\newcommand{\osc}[1]{\mathbf{#1}}

\newcommand{\Tr}{\mathop{\mathrm{Tr}}}
\newcommand{\STr}{\mathop{\mathrm{STr}}}

\newcommand{\Integers}{\mathbb{Z}}
\newcommand{\Complex}{\mathbb{C}}
\newcommand{\Reals}{\mathbb{R}}

\ifx\genfrac\sdflkaj\else\fi
\newcommand{\sfrac}[2]{{\textstyle\frac{#1}{#2}}}
\newcommand{\half}{\sfrac{1}{2}}
\newcommand{\ihalf}{\sfrac{i}{2}}
\newcommand{\quarter}{\sfrac{1}{4}}

\newcommand{\indup}[1]{_{\mathrm{#1}}}

\newcommand{\rep}[1]{{\mathbf{#1}}}
\newcommand{\matr}[2]{\left(\begin{array}{#1}#2\end{array}\right)}

\newcommand{\bigbrk}[1]{\bigl(#1\bigr)}

\newcommand{\bigvev}[1]{\bigl\langle#1\bigr\rangle}

\newcommand{\comm}[2]{[#1,#2]}

\newcommand{\acomm}[2]{\{#1,#2\}}

\newcommand{\gcomm}[2]{\mathopen{[}#1,#2\mathclose{\}}}
\newcommand{\biggcomm}[2]{\bigl[#1,#2\bigr\}}

\newcommand{\abs}[1]{|#1|}

\newcommand{\bigsetspec}[2]{\bigl\{#1\big|#2\bigr\}}
\newcommand{\state}[1]{\mathopen{|}#1\mathclose{\rangle}}

\newcommand{\alg}[1]{\mathfrak{#1}}
\newcommand{\grp}[1]{\mathrm{#1}}

\newcommand{\nln}{\nonumber\\}

\newcommand{\earel}[1]{\mathrel{}&\hspace{-2\arraycolsep}#1\hspace{-2\arraycolsep}&\mathrel{}}
\newcommand{\eq}{\earel{=}}

\def\[{\begin{equation}}
\def\]{\end{equation}}
\def\<{\begin{eqnarray}}
\def\>{\end{eqnarray}}

\makeatletter
\def\mr@ignsp#1 {\ifx\:#1\@empty\else #1\expandafter\mr@ignsp\fi}%
\newcommand{\multiref}[1]{\begingroup
\xdef\mr@no@sparg{\expandafter\mr@ignsp#1 \: }%
\def\mr@comma{}%
\@for\mr@refs:=\mr@no@sparg\do{\mr@comma\def\mr@comma{,}\ref{\mr@refs}}%
\endgroup}
\makeatother

\newcommand{\hypref}[2]{\ifx\href\asklfhas #2\else\href{#1}{#2}\fi}

\newcommand{\secref}[1]{Sec.~\multiref{#1}}

\newcommand{\figref}[1]{Fig.~\multiref{#1}}
\renewcommand{\eqref}[1]{(\multiref{#1})}

\makeatletter
\newlength{\apb@width}
\newcommand{\autoparbox}[2][c]{\settowidth{\apb@width}{#2}\parbox[#1]{\apb@width}{#2}}
\newcommand{\includegraphicsbox}[2][]{\autoparbox{\includegraphics[#1]{#2}}}
\makeatother

\providecommand{\href}[2]{#2}
\providecommand{\arxivlink}[1]{\href{http://arxiv.org/abs/#1}{arxiv:#1}}

\begin{document}


\thispagestyle{empty}
\phantomsection
\addcontentsline{toc}{section}{Title}

\begin{flushright}\footnotesize%
\texttt{AEI-2010-174},
\texttt{\arxivlink{1012.4004}}\\
overview article: \texttt{\arxivlink{1012.3982}}%
\vspace{1em}%
\end{flushright}

\begingroup\parindent0pt
\begingroup\bfseries\ifarxiv\Large\else\LARGE\fi
\hypersetup{pdftitle={Review of AdS/CFT Integrability, Chapter VI.1: Superconformal Symmetry}}%
Review of AdS/CFT Integrability, Chapter VI.1:\\
Superconformal Symmetry
\par\endgroup
\vspace{1.5em}
\begingroup\ifarxiv\scshape\else\large\fi%
\hypersetup{pdfauthor={Niklas Beisert}}%
Niklas Beisert
\par\endgroup
\vspace{1em}
\begingroup\itshape
Max-Planck-Institut f\"ur Gravitationsphysik,
Albert-Einstein-Institut\\
Am M\"uhlenberg 1,
14476 Potsdam,
Germany
\par\endgroup
\vspace{1em}
\begingroup\ttfamily
nbeisert@aei.mpg.de
\par\endgroup
\vspace{1.0em}
\endgroup

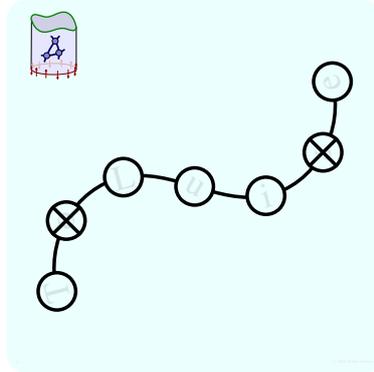
\begin{center}
\includegraphics[width=5cm]{TitleVI1.mps}
\vspace{1.0em}
\end{center}

\paragraph{Abstract:}
Aspects of the $D=4$, $\superN=4$ superconformal symmetry
relevant to the AdS/CFT duality and integrability are reviewed.
These include the Lie superalgebra $\alg{psu}(2,2|4)$,
its representations, conformal transformations and
correlation functions in $\superN=4$ super Yang--Mills theory
as well as an illustration of the $AdS_5\times S^5$
superspace on which the dual string theory is formulated.

\ifarxiv\else
\fi
\hypersetup{pdfsubject={MSC (2010): 17B10, 81T13, 81T30, 81T60}}%

\ifarxiv\else
\fi
\hypersetup{pdfkeywords={PSU(2,2|4), representations, correlation functions, AdS5xS5}}%

\newpage


\section{Introduction}

The AdS/CFT correspondence predicts the
exact equivalence of $\superN=4$ super Yang--Mills (SYM) theory 
with IIB superstrings propagating on the $AdS_5\times S^5$ background.
One of the immediate checks is that the two models have 
coincident global symmetries: $\superN=4$ superconformal symmetry on the one hand
and the isometries of the $AdS_5\times S^5$ superspace on the other 
are both given by the Lie supergroup $\widetilde{\grp{PSU}}(2,2|4)$ 
or its algebra $\alg{psu}(2,2|4)$.

Symmetry serves as an important organising principle ---
e.g.\ for objects with similar properties --- and 
leads to structural constraints --- e.g.\ for correlation functions. 
Furthermore, supersymmetry often implies that selected
quantities are protected from receiving quantum corrections. 
Two famous examples are 
the exact quantum conformal symmetry of $\superN=4$ SYM
due to absence of a beta-function
\cite{Sohnius:1981sn,Mandelstam:1982cb,Brink:1982pd,Howe:1983sr}
and the exactness of correlators for certain BPS operators
in agreement with a prediction of the AdS/CFT duality,%
see the review \cite{D'Hoker:2002aw}.
Nevertheless, agreement of the symmetry groups is far from sufficient 
to prove an exact duality.%
\footnote{In this case the large amount of
(super)symmetry at least makes both constituent models essentially unique 
and exceptional, which may be viewed as a hint towards 
the validity of the correspondence.}
To verify the AdS/CFT conjecture one therefore needs tests 
involving dynamical quantities which are not protected by the symmetry. 
Much of the activity concerning AdS/CFT integrability
is devoted to such tests. 
Making use of superconformal symmetry has helped
the progress at various stages.

The present paper reviews some aspects of the Lie superalgebra $\alg{psu}(2,2|4)$
relevant to AdS/CFT integrability.
The presented facts are by no means restricted to integrability; 
they were known long before AdS/CFT integrability was discovered,
and little progress was made in connection with the latter.
Nevertheless, many results in AdS/CFT integrability are based
on a good knowledge of $\alg{psu}(2,2|4)$. 
This paper therefore serves a different purpose than the other chapters
of the review collection \cite{chapIntro}: 
It is not so much a review of one particular aspect of AdS/CFT integrability, 
but should be viewed as a reference guide to key concepts concerning 
the underlying global symmetry. 

This paper is split into three parts:
In \secref{sec:Algebra} we shall review 
purely algebraic aspects of $\alg{psu}(2,2|4)$
such as the algebra itself as well as some essential representation theory.
In \secref{sec:Conformal} we apply it to local operators in $\superN=4$ SYM
and their correlation functions.
In \secref{sec:Isometries} we discuss the $AdS_5\times S^5$ 
background on which superstrings can propagate and which is
a particular coset of $\widetilde{\grp{PSU}}(2,2|4)$.

\section{The \texorpdfstring{$\alg{psu}(2,2|4)$}{psu(2,2|4)} Algebra}
\label{sec:Algebra}

\paragraph{Definition.}

The algebra $\alg{psu}(2,2|4)$ is a real Lie superalgebra
of (even$|$odd) dimension $30|32$,
see e.g.\ \cite{Cornwell:1989bx,Frappat:2000aa,Frappat:1996pb}. 
In order to define it, it is convenient to start 
with complex $4|4$-dimensional 
square supermatrices
\[\label{eq:Xmat}
X=\matr{c|c}{A&B\\\hline C&D}.
\]
Each block $A,B,C,D$ is a $4\times 4$ matrix of (non-Gra{\ss}mannian) complex numbers.
The blocks $A,D$ are considered \emph{even} and $B,C$ \emph{odd}.
The Lie superalgebra $\alg{gl}(4|4,\Complex)$ 
is the $32|32$-dimensional vector space of these supermatrices.
Its graded Lie bracket $\gcomm{\cdot}{\cdot}$ is defined as the graded commutator of supermatrices
(in the following $Y$ is the analog of $X$ in \eqref{eq:Xmat} with blocks $E,F,G,H$)
\[
\gcomm{X}{Y}=XY-(-1)^{XY}YX:=\matr{c|c}{AE+BG-EA+FC&AF+BH-EB-FD\\\hline CE+DG-GA-HC&CF+DH+GB-HD}.
\]
It differs from a conventional commutator through the signs for the
odd-odd products $FC$ and $GB$.
It also satisfies a graded Jacobi-identity
\[
(-1)^{XZ}\biggcomm{\gcomm{X}{Y}}{Z}
+(-1)^{YX}\biggcomm{\gcomm{Y}{Z}}{X}
+(-1)^{ZY}\biggcomm{\gcomm{Z}{X}}{Y}
=0.
\]
This algebra is not simple, it has non-trivial ideals:
One is related to the supertrace $\STr X:=\Tr A-\Tr D$
which is zero for graded commutators $\STr\gcomm{X}{Y}=0$.
Demanding that $\STr X=0$ thus removes a derivation from $\alg{gl}(4|4,\Complex)$, 
and restricts it to the subalgebra $\alg{sl}(4|4,\Complex)$.
Furthermore, the identity supermatrix $1$ 
commutes with all other matrices, $\gcomm{1}{X}=0$.
Hence it generates the centre and can be projected
out from $\alg{gl}(4|4,\Complex)$
yielding $\alg{pgl}(4|4,\Complex)$.
The combination of restriction and projection
is the $30|32$-dimensional complex
Lie superalgebra $\alg{psl}(4|4,\Complex)$.%
\footnote{It is not possible to restrict to $\Tr A=\Tr D=0$
because the graded commutator does \emph{not close}
onto such supermatrices; 
The centre proportional to the unit supermatrix can only be \emph{projected} out
or removed by redefining the graded commutator accordingly.}

\paragraph{Real Form.}

To restrict to the real form $\alg{psu}(2,2|4)$
one imposes a hermiticity condition
on the supermatrices
\[
\matr{c|c}{A&B\\\hline C&D}
=
\matr{c|c}{HA^\dagger H^{-1}&-iHC^\dagger\\\hline -iB^\dagger H^{-1}&D^\dagger},
\]
where $H$ is a hermitian matrix of signature $(2,2)$.
There are two natural choices for $H$:
In the first, $H$ is diagonal, written in terms of $2\times 2$ blocks
(`$+$'/`$-$' denotes the $2\times2$ positive/negative identity matrix; $X'$ is a reordering of rows and columns to be explained)
\[
\label{eq:PSUAdS}
H=\matr{cc}{+&0\\0&-},
\quad
X=\matr{cc|c}{
M_1&iN&-iQ_1\\
i\bar N&M_2&+iQ_2\\\hline
\bar Q_1&\bar Q_2\vphantom{\hat Q}&R
},
\quad
X'=\matr{c|c|c}{
M_1&-iQ_1&iN\\\hline
\bar Q_1&R&\bar Q_2\vphantom{\hat Q}\\\hline
i\bar N&+iQ_2&M_2
}.
\]
Here the hermitian blocks $M_1$ and $M_2$ generate the maximal compact 
subalgebra 
$\alg{su}(2)\oplus\alg{su}(2)\oplus\alg{u}(1)
=\alg{so}(4)\oplus\alg{so}(2)$
of $\alg{su}(2,2)=\alg{so}(4,2)$.
This choice is useful in the context of the $AdS_5$ spacetime,
cf.\ \secref{sec:Isometries},
and for unitary representations.
Equivalently one can choose an off-diagonal $H$
\[
\label{eq:PSUMinkowski}
H=\matr{cc}{0&+\\+&0},\quad
X=\matr{cc|c}{
L&P&-iQ\\
K&\bar L\vphantom{\hat Q}&-i\bar S\\\hline
S&\bar Q\vphantom{\hat Q}&R
},\quad
X'=\matr{c|c|c}{
L&-iQ&P\\\hline
S&R&\bar Q\vphantom{\hat Q}\\\hline
K&-i\bar S\vphantom{\hat Q}&\bar L
}.
\]
Now the hermitian conjugate blocks $L,\bar L$ in $X$ generate the 
Lorentz and scaling transformations in
$\alg{sl}(2,\Complex)\oplus\alg{gl}(1)=\alg{so}(3,1)\oplus\alg{so}(1,1)$.
Obviously, this choice is adapted to four-dimensional Minkowski space,
see \secref{sec:Conformal}.
In the context of the real form $\alg{psu}(2,2|4)$ it is often convenient to reorder 
the $2,2|4$ rows and columns, and move one of the $2$'s past the $4$.
The supermatrix $X$ reordered in $2|4|2$-block form 
is displayed in \eqref{eq:PSUAdS,eq:PSUMinkowski} as $X'$.
From now on we shall use exclusively the $2|4|2$-grading.

\paragraph{Simple Generators.}

A useful presentation of Lie algebras,
which is frequently encountered in the solution of integrable systems,
is through $r$ triplets of \emph{simple} (raising, Cartan and lowering) generators $\gen{E}_k,\gen{H}_k,\gen{F}_k$,
($r$ is the rank of the algebra),
see e.g.\ \cite{Cornwell:1997ke}.
For the Lie algebras $\alg{sl}(n)$ the elements $\gen{E}_k,\gen{H}_k,\gen{F}_k$
with $k=1,\ldots,n-1$, generate the three main diagonals 
$X_{k,k+1},X_{k,k}-X_{k+1,k+1},X_{k+1,k}$.
The remaining elements are obtained by repeated Lie brackets,
e.g.\ $\gcomm{\gen{E}_k}{\gen{E}_{k+1}}$ generates $X_{k,k+2}$.
Evidently, the algebra generated by arbitrary repeated brackets
is enormous and needs to be reduced by certain relations.
To that end, 
the simple generators satisfy a set of Chevalley--Serre relations
which encode all the information on the specific Lie algebra,
$\alg{sl}(n)$,
in a condensed form 
\[
\begin{array}[b]{c}
\gcomm{\gen{H}_j}{\gen{E}_k}=+A_{jk}\gen{E}_k,\qquad
\gcomm{\gen{H}_j}{\gen{F}_k}=-A_{jk}\gen{F}_k,\qquad
\gcomm{\gen{E}_j}{\gen{F}_k}=\delta_{jk}\gen{H}_k,
\\[1ex]
\biggcomm{\gcomm{\gen{E}_k}{\gen{E}_{k\pm 1}}}{\gen{E}_{k\pm 1}}=\biggcomm{\gcomm{\gen{F}_k}{\gen{F}_{k\pm 1}}}{\gen{F}_{k\pm 1}}=0,
\\[1ex]
\gcomm{\gen{E}_j}{\gen{E}_k}=\gcomm{\gen{F}_j}{\gen{F}_k}=0\mbox{ for }|j-k|>1.
\end{array}
\]
Here $A_{j,k}$ is the Cartan matrix; 
for $\alg{sl}(n)$ the three main diagonals take the values $-1,+2,-1$
while the other elements are zero.
For a superalgebra $\alg{sl}(n|m)$ the definition is similar;
the main difference is that some of the raising and lowering elements are odd. 
For an odd $\gen{E}_k$ ($k=2,6$ in our case) one has to replace the
relation $\gcomm{\gcomm{\gen{E}_{k\pm 1}}{\gen{E}_k}}{\gen{E}_k}=0$
by two new ones%
\footnote{It turns out that in gauge/string applications 
the latter relations are dropped, see e.g.\ \cite{Beisert:2003ys}.
The new generators
$\gen{G}_k\sim\gcomm{\gcomm{\gen{E}_{k-1}}{\gen{E}_k}}{\gcomm{\gen{E}_{k+1}}{\gen{E}_k}}$
(similarly for $\gen{F}_k$)
are part of an ideal of a substantially bigger algebra. 
The ideal generates gauge transformations
acting as the constraint $\gen{G}_k\simeq 0$ for physical states.}
\[
\gcomm{\gen{E}_k}{\gen{E}_k}=0,
\qquad
\biggcomm{\gcomm{\gen{E}_{k-1}}{\gen{E}_k}}{\gcomm{\gen{E}_{k+1}}{\gen{E}_k}}=0,
\]
and similarly for $\gen{F}_k$. Furthermore for this $k$ 
two Cartan matrix elements are modified: $A_{k,k}=0$ and $A_{k,k+1}=+1$.
Cartan matrices and Chevalley--Serre relations are
often displayed in the form of Dynkin diagrams. 
Two Dynkin diagrams for $\alg{su}(2,2|4)$ are
displayed in \figref{fig:Dynkin}:
\begin{figure}\centering
\includegraphics{FigDynkinOOOXOOO.mps}
\qquad
\includegraphics{FigDynkinOXOOOXO.mps}
\caption{Two Dynkin diagrams for $\alg{sl}(4|4)=\alg{sl}(2|4|2)$.}
\label{fig:Dynkin}
\end{figure}
Dots correspond to simple generators $\gen{E}_k,\gen{H}_k,\gen{F}_k$;
crossed dots indicate odd generators $\gen{E}_k,\gen{F}_k$.
Links stand for non-trivial relations between the corresponding simple generators
and non-trivial Cartan matrix elements.
If two dots $j$ and $k$ are unlinked, 
the generators $\gen{E}_k,\gen{F}_k,\gen{H}_k$ and $\gen{E}_j,\gen{F}_j,\gen{H}_j$ commute and $A_{jk}=0$.
Although the two Dynkin diagrams lead to quite different relations,
they describe the same algebra. 
The point is that for Lie superalgebras
there commonly exist inequivalent choices for the set of simple generators.
The two diagrams correspond to the two grading assignments $4|4$ and $2|4|2$
for the rows and columns of a supermatrix, 
cf.\ $X$ vs.\ $X'$ in \eqref{eq:PSUAdS,eq:PSUMinkowski}.

\paragraph{Unitary Representations.}

In physical models, multiplets of states 
transform under \emph{unitary} representations
of the symmetry algebra.
Let us therefore review unitary representations of $\alg{psu}(2,2|4)$
\cite{Dobrev:1985qv}.
As the (bosonic part of the) superalgebra is non-compact, 
unitary representations are necessarily infinite-dimensional.
An important class of unitary representations 
are the \emph{lowest-weight} (equivalently highest-weight) representations.
Under the maximal compact subalgebra 
$\alg{su}(2)\oplus\alg{su}(2)\oplus\alg{su}(4)\oplus\alg{u}(1)$
such representations decompose into 
(infinitely many) finite-dimensional irreps,
one of which is defined as the lowest.
All states corresponding to this lowest irrep are annihilated
by the lowering generators associated to the lower 
triangular blocks $\bar Q_1,Q_2,\bar N$ of $X'$ in \eqref{eq:PSUAdS}.
The states of the higher irreps arise from the repeated 
action of the raising generators associated to the upper 
triangular blocks $Q_1,\bar Q_2,N$ of $X'$.

\begin{figure}\centering
\includegraphics{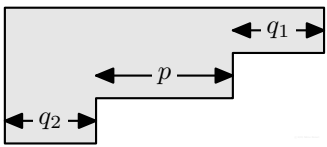}
\caption{Young diagram corresponding to $\alg{sl}(4)$ representation with Dynkin labels $[q_1,p,q_2]$:
A single block corresponds to a fundamental representation, rows and columns
correspond to symmetrisation and antisymmetrisation.}
\label{fig:Young}
\end{figure}

Lowest-weight unitary representations of $\alg{psu}(2,2|4)$ are thus 
specified by an irrep under the maximal compact subalgebra
$\alg{su}(2)\oplus\alg{su}(2)\oplus\alg{su}(4)\oplus\alg{u}(1)$.
Irreps of the two $\alg{su}(2)$'s are specified by their non-negative half-integer spin $\half s_{1,2}$
or equivalently by the non-negative integer \emph{Dynkin labels} $[s_1]$ and $[s_2]$.
Analogously, irreps of $\alg{su}(4)$ are specified through three 
non-negative integer Dynkin labels $[q_1,p,q_2]$.
An alternative description uses a Young diagram with no more than three rows,
see \figref{fig:Young}, cf.\ \cite{Sternberg:1994tw}
Finally, a $\alg{u}(1)$ irrep is specified through a number $E$.
Here there is a subtlety: 
The abelian algebra $\alg{u}(1)=\Reals$ can either generate the 
compact group $\grp{U}(1)$ or the non-compact additive group $\Reals$.
For a compact group $E$ is restricted to an integer
whereas a non-compact group merely requires $E$ to be real.
The supergroup $\grp{PSU}(2,2|4)$ contains the 
compact version and hence the spectrum of $E$ is discrete.
However, $\grp{PSU}(2,2|4)$ has a non-trivial
\emph{universal cover} $\widetilde{\grp{PSU}}(2,2|4)$ where the abelian subgroup becomes non-compact. 
It is this universal cover which has applications to physics,
and consequently we shall allow continuous values for $E$.

Altogether, a unitary representation is specified by the Dynkin labels
$[s_1]$, $[s_2]$, $[q_1,p,q_2]$ and the number $E$. 
These combine into $\alg{su}(2,2|4)$ Dynkin labels: 
\[
[s_1;r_1;q_1,p,q_2;r_2;s_2],
\qquad
r_k=\half E+\half s_k-\sfrac{3}{4}q_k-\half p-\sfrac{1}{4}q_{3-k}.
\]
Finally, we should note that the value of $E$ must be above a certain bound
which is most conveniently expressed in terms of the $r_k$
\[\label{eq:UniConstr}
r_k\geq 1+s_k\quad \mbox{or}\quad r_k=s_k=0.
\]
If one of the bounds for the first condition is saturated
or one of the second conditions is satisfied,
the representation is called \emph{atypical} or short.
In this case certain combinations of the raising generators
annihilate the lowest-weight state.
Otherwise there are no additional restrictions on the representation 
of the raising generators, and the representation is called \emph{typical} or long.

\section{Superconformal Symmetry in \texorpdfstring{$\superN=4$}{N=4} SYM}
\label{sec:Conformal}

For $\superN=4$ supersymmetric gauge theory 
on four-dimensional Minkowski space 
the super-Poincar\'e algebra
extends to the superconformal algebra $\alg{psu}(2,2|4)$.
In the following we shall discuss the representation theory of
$\alg{psu}(2,2|4)$ related to this gauge theory,
see also \cite{D'Hoker:2002aw} for an extended review.

\paragraph{Conformal Transformations.}

Conformal transformations preserve the metric up to a 
local rescaling of distances.
In four-dimensional Minkowski space 
conformal symmetry is based on the Lie algebra $\alg{so}(4,2)=\alg{su}(2,2)$.
It contains the $\alg{sl}(2,\Complex)$ Lorentz rotations $\gen{L},\gen{\bar L}$
and translations $\gen{P}$ which form the Poincar\'e algebra. 
In addition, there are the dilatation $\gen{D}$ and the 
conformal boosts $\gen{K}$. 
The extension to the superconformal algebra 
consists of the internal $\alg{su}(4)$ rotations $\gen{R}$,
the supertranslations $\gen{Q},\gen{\bar Q}$ as well as
the superconformal boosts $\gen{S},\gen{\bar S}$.
These generators correspond to the submatrices in \eqref{eq:PSUMinkowski}.

The conformal generators $\gen{P},\gen{D},\gen{K}$ 
act on the coordinates $x^\mu$ of Minkowski space 
with metric $\eta^{\mu\nu}$ as
\[\label{eq:ConfCoord}
\gen{P}_\mu x^\nu =i\delta_\mu^\nu,
\qquad
\gen{D} x^\mu =ix^\mu,
\qquad
\gen{K}^\mu x^\nu =ix^\mu x^\nu-\ihalf\eta^{\mu\nu}x\mathord{\cdot} x.
\]
The action of the odd generators 
is rather complicated and requires the introduction of
fermionic coordinates; we refrain from spelling out the explicit form.
Fields on Minkowski space transform according to the above rules, 
but in addition they have intrinsic transformation properties
such as \emph{spin} and \emph{conformal dimension}.
For example, the conformal representation on a scalar primary field $\Phi(x)$
of dimension $d$ reads
\[\label{eq:ConfField}
\gen{P}_\mu \Phi=i\partial_\mu\Phi,
\quad
\gen{D} \Phi = id\Phi +ix\mathord{\cdot}\partial\Phi,
\quad
\gen{K}^\mu \Phi =idx^\mu\Phi+ix^\mu x\mathord{\cdot} \partial\Phi-\ihalf x\mathord{\cdot} x \partial^\mu\Phi.
\]
Representations for fields with spin are slightly more complicated,
and representations of the complete superconformal algebra
suggest the use of fields on superspace. Both of these
aspects will not be considered explicitly.

\paragraph{Correlators.}

The power of conformal symmetry is that it constrains correlation functions 
in a conformal quantum field theory, see e.g.\ \cite{Mack:1969rr}.
In particular, the spacetime dependence of two- and three-point 
functions is fully determined
\<
\bigvev{\Phi_1(x)\,\Phi_2(y)}\eq\frac{N}{|x-y|^{2d}}\,,\qquad
(\mbox{requires }d_1=d_2=d),
\nln
\bigvev{\Phi_1(x)\,\Phi_2(y)\,\Phi_3(z)}\eq
\frac{C_{123}}{|x-y|^{d_1+d_2-d_3}|y-z|^{d_2+d_3-d_1}|z-x|^{d_3+d_1-d_2}}\,.
\>
Indices for fields with spin are typically contracted
with suitable tensors, e.g.\ $I^{\mu\nu}=\eta^{\mu\nu}-2(x-y)^\mu(x-y)^\nu/(x-y)^2$.
The reason for complete determination is that 
any three points can be mapped to any other three points
by conformal transformations. 
The value of the correlator at one configuration of three points
thus determines the value of the correlator at any other configuration.
For four or more points there exist conformally invariant cross ratios,
e.g.\ $\abs{x_{12}}\abs{x_{34}}/\abs{x_{13}}\abs{x_{24}}$,
on which the correlation functions can depend without constraints. 
Note that there exist superconformal cross ratios of the fermionic coordinates
already for three points in superspace.%
\footnote{The number of invariants is related to
the dimension of the group, 
the dimension of the stabiliser 
and the number of coordinates.
E.g., three points in superspace have 48 fermionic coordinates,
but the group has only 32. 
Hence there should be 16 invariant combinations of
fermionic coordinates.}

The above constraints on correlators hold 
for all fields which have well-defined transformation properties
under superconformal symmetry. 
This includes the fundamental fields (to some extent),
but more importantly also \emph{composite local operators}. 
The latter are local products of the fundamental fields 
and their derivatives.
In the free field theory, they transform in tensor products 
of the fundamental field representation.
Let us therefore discuss the superconformal representations
that come to use.

\paragraph{Fundamental Field Representation.}

Consider first a scalar field $\Phi$ in four dimensions.
In the free theory $\Phi$ obeys the conformal transformation 
rules \eqref{eq:ConfField} with $d=1$. 
For a local operator we shall need $\Phi$ and its derivatives
at the point $x$ which for convenience we assume to be 
the origin of spacetime $x=0$. In other words, 
we represent $\Phi(x)$ through its Taylor series around $x=0$
\[
\Phi(x)=\Phi(0)+x^\mu\partial_\mu\Phi(0)+\half x^\mu x^\nu\partial_\mu\partial_\nu\Phi(0)+\ldots\,.
\]
We can now see that the conformal representation \eqref{eq:ConfField}
acts on these Taylor components (we drop the argument $x=0$):
\[\label{eq:ConfTaylor}
\begin{array}[b]{rclcrclcrcl}
\gen{P}_\mu \Phi\eq i\partial_\mu\Phi,
&&
\gen{D} \Phi \eq  id\Phi ,
&&
\gen{K}^\mu \Phi \eq 0,
\\[0.6ex]
\gen{P}_\mu \partial_\rho\Phi\eq i\partial_\rho\partial_\mu\Phi,
&&
\gen{D} \partial_\rho\Phi \eq i(d+1)\partial_\rho\Phi ,
&&
\gen{K}^\mu \partial_\rho\Phi \eq id\delta_\rho^\mu\Phi,
\\[0.6ex]
\earel{} && \earel{}\ldots. && \earel{}
\end{array}
\]
This is a lowest-weight representation, where $\gen{K}$ serves as the lowering
generator to annihilate the primary field $\Phi$.
The raising generator $\gen{P}$ is used to 
access the descendants
$\partial_\mu\Phi$, $\partial_\mu\partial_\nu\Phi$, \ldots, while
$\gen{D}$ essentially measures the number of derivatives.

There is one noteworthy peculiarity of the boost acting on 
$\partial_\rho\partial_\sigma\Phi$
\[
\gen{K}^\mu  \partial_\rho\partial_\sigma \Phi =
i(d+1)  \delta_\sigma^\mu\partial_\rho\Phi
+i(d+1)  \delta_\rho^\mu\partial_\sigma\Phi
-i     \eta_{\rho\sigma} \partial^\mu\Phi.
\]
When acting on the D'Alembertian derivative $\partial\mathord{\cdot}\partial\Phi$
one obtains $2i(d-1)\partial^\mu\Phi$
which vanishes precisely for the physical scaling dimension $d=1$.
This means that the lowest-weight representation is reducible,
and we should divide out a subrepresentation by 
imposing the free equation of motion $\partial\mathord{\cdot}\partial\Phi=0$.

The equation of motion implies the absence of certain components
in the Taylor expansion. The enumeration of non-trivial components
is most transparent when using pairs of spinor indices $\beta\dot\alpha$
instead of the vector indices $\mu$. 
Now a trace $\eta^{\mu\nu}$ is replaced by a
pair of antisymmetric $\alg{sl}(2,\Complex)$ invariants
$\varepsilon^{\beta\delta}\varepsilon^{\dot\alpha\dot\gamma}$.
For any pair of derivatives we can thus exclude antisymmetrisation in both 
pairs of spinor indices by virtue of the equations of motion. 
Furthermore, due to the commutative nature of derivatives, 
antisymmetrisation in just one pair of spinor indices is also zero.
Effectively it means that all spinor indices of either kind 
must be fully symmetrised. 
Such symmetrisation is automatic for states of a four-dimensional \emph{harmonic oscillator}:
We can replace
\[
\partial_{\beta\dot\alpha}
\partial_{\delta\dot\gamma}\ldots
\Phi
\simeq
\osc{\bar a}_{\beta}
\osc{\bar a}_{\delta}
\ldots
\osc{\bar b}_{\dot\alpha}
\osc{\bar b}_{\dot\gamma}
\ldots
\state{0},
\]
where the algebra of creation and annihilation operators 
is defined through the non-trivial commutation relations
\[
\comm{\osc{a}^\alpha}{\osc{\bar a}_\gamma}=i\delta^\alpha_\gamma,\qquad
\comm{\osc{b}^{\dot\alpha}}{\osc{\bar b}_{\dot\gamma}}=i\delta^{\dot\alpha}_{\dot\gamma},\qquad
\acomm{\osc{c}^{a}}{\osc{\bar c}_{c}}=\delta^{a}_{c}.
\]
Here we have added a set of four fermionic oscillators $\osc{c}$
which make the generalisation to all fields of $\superN=4$ straight-forward: 
States have up to four excitations of $\osc{\bar c}$ transforming in 
the $\alg{su}(4)$ representations $\rep{1},\rep{4},\rep{6},\rep{\bar{4}},\rep{1}$,
respectively.
This matches precisely with the representations of
the chiral part of the gauge field strength $\Gamma_{\alpha\gamma}$, 
the chiral fermions $\Psi_{\alpha c}$,
the scalars $\Phi_{ac}$,
the antichiral fermions $\bar\Psi^c_{\dot\alpha}$
and the antichiral field strength $\bar\Gamma_{\dot\alpha\dot\gamma}$.%
\footnote{
The field strength $\Gamma_{\mu\nu}$
with antisymmetric vector indices
decomposes into two complex conjugate fields
$\Gamma_{\alpha\gamma}$ and $\bar\Gamma_{\dot \alpha\dot \gamma}$
with symmetric spinor indices.
Similarly, a real $\alg{so}(6)$ vector of fields $\Phi_m$ 
is equivalent to a field $\Phi_{ac}$
with antisymmetric $\alg{su}(4)$ indices
and reality condition
$\Phi_{ac}=\half\varepsilon_{acbd}\bar\Phi^{bd}$.
}
Altogether, for every state of the supersymmetric oscillator,
subject to the constraint
\[\label{eq:OscConstr}
\osc{N_a}-\osc{N_b}+\osc{N_c}\simeq 2,
\]
there is exactly one Taylor component of the on-shell fundamental fields 
of $\superN=4$ SYM \cite{Gunaydin:1984fk}. 
The excitation number operators are defined as
$
\osc{N_a}:=-i \osc{\bar a}_\alpha \osc{a}^\alpha$, 
$\osc{N_b}:=-i \osc{\bar b}_{\dot\alpha} \osc{b}^{\dot\alpha}$,
$\osc{N_c}:=\osc{\bar c}_a \osc{c}^a$.

The oscillator basis is also particularly convenient 
for the superconformal algebra: All the generators are represented
through bilinears in the oscillators:
\[\begin{array}[b]{rclcrcl}
\gen{L}^\alpha{}_\gamma\earel{\simeq}
\osc{\bar a}_\gamma\osc{a}^\alpha
-\half\delta_\gamma^\alpha \osc{\bar a}_\epsilon\osc{a}^\epsilon,
&&
\gen{R}^a{}_c\earel{\simeq}
\osc{\bar c}_c\osc{c}^a
-\quarter\delta_c^a \osc{\bar c}_e\osc{c}^e,
\\[0.6ex]
\gen{\bar L}_{\dot\gamma}{}^{\dot\alpha}\earel{\simeq}
\osc{b}^{\dot\alpha}\osc{\bar b}_{\dot\gamma}
-\half\delta_{\dot\gamma}^{\dot\alpha} \osc{b}^{\dot\epsilon}\osc{\bar b}_{\dot\epsilon},
&&
\gen{D}\earel{\simeq}
\half \osc{\bar a}_\alpha \osc{a}^\alpha
+\half  \osc{b}^{\dot\alpha}\osc{\bar b}_{\dot\alpha},
\\[0.6ex]
\gen{P}_{\gamma\dot\alpha}\earel{\simeq}
\osc{\bar a}_{\gamma}\osc{\bar b}_{\dot\alpha},
&&
\gen{K}^{\gamma\dot\alpha}\earel{\simeq}
\osc{b}^{\dot\alpha}\osc{a}^{\gamma},
\\[0.6ex]
\gen{Q}^a_{\gamma}\earel{\simeq}
\osc{\bar a}_{\gamma}\osc{c}^{a},
&&
\gen{S}_a^{\gamma}\earel{\simeq}
\osc{\bar c}_{a}\osc{a}^{\gamma},
\\[0.6ex]
\gen{\bar Q}_{\dot\gamma a}\earel{\simeq}
\osc{\bar c}_{a}\osc{\bar b}_{\dot\gamma},
&&
\gen{\bar S}^{\dot\gamma a}\earel{\simeq}
\osc{b}^{\dot\gamma}\osc{c}^{a}.
\end{array}\]
These satisfy the $\alg{psu}(2,2|4)$ algebra
along with its reality conditions provided that
\[
(\osc{\bar a}_{\alpha})^\dagger=\osc{\bar b}_{\dot\alpha},
\qquad
(\osc{a}^{\alpha})^\dagger=\osc{b}^{\dot\alpha} ,
\qquad
(\osc{\bar c}_a)^\dagger=\osc{c}^{a} .
\]
The algebra extends to $\alg{u}(2,2|4)$ by introducing 
a derivation $\gen{B}\simeq\osc{\bar c}_a \osc{c}^a$ and a central charge 
$\gen{C}\simeq-i \osc{\bar a}_\alpha \osc{a}^\alpha+i \osc{b}^{\dot\alpha}\osc{\bar b}_{\dot\alpha}+\osc{\bar c}_a \osc{c}^a$.
The constraint \eqref{eq:OscConstr} is equivalent to the 
vanishing of the central charge, hence the above form
a consistent representation of $\alg{psu}(2,2|4)$.

Note that the above construction remains applicable to the interacting theory
for the sake of enumerating local composite operators: 
The r.h.s.\ of the equation of motion
$\partial\mathord{\cdot}\partial\Phi=\ldots$ is not zero, 
but it is a product of fields which is 
already accounted for in the basis of local operators.
Furthermore, to maintain proper gauge transformation properties, 
partial derivatives should be replaced by their covariant counterparts. 
Consequently, antisymmetries of derivatives are no longer excluded.
They lead to commutators with the field strength, 
which are again accounted for in the basis of local operators.
The only change in the quantum theory is
that the representation on composite operators is deformed in a specific way,
see the chapters \cite{chapChain,chapHigher,chapLR}.
For example, the scaling dimensions of composite operators 
generically receive continuous quantum corrections.

\paragraph{Composite Operator Multiplets.}

Composite operators are local products of the fundamental fields
and hence they transform in tensor products of the above representation.
Tensor products of lowest-weight representations typically decompose
into sums of lowest-weight representations. 
Thus composite operators form multiplets each of which has a primary field. 

\begin{figure}
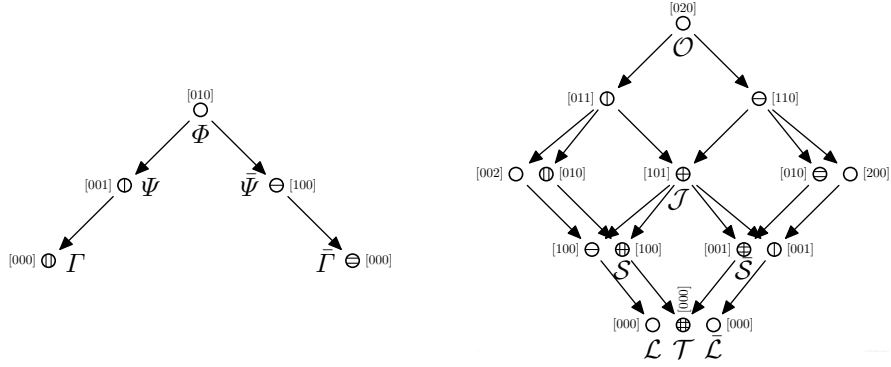
\centering
\includegraphicsbox{FigMultiField.mps}
\qquad
\includegraphicsbox{FigMultiCurr.mps}
\caption{Field multiplet $[0;0;0,1,0;0;0]$
(top component $\Phi$ at $d=1$)
and current multiplet $[0;0;0,2,0;0;0]$
(top component $\Op{O}$ at $d=2$).
Each dot corresponds to a field of $\alg{su}(2,2)\oplus\alg{su}(4)$:
The two $\alg{su}(2)$ spins are indicated by horizontal/vertical bars,
while the $\alg{su}(4)$ representation is indicated through 
Dynkin labels.
SW/SE arrows correspond to 
the action of the Poincar\'e supercharges $\gen{Q}$/$\bar{\gen{Q}}$.}
\label{fig:Multi}
\end{figure}

The simplest non-trivial local operator is
a traceless combination of two scalars%
\footnote{In a gauge theory one should pick a gauge-invariant combination.}
$\Op{O}_{mn}=\Phi_m\Phi_n-\frac{1}{6}\delta_{mn}\Phi_p\Phi_p$
transforming as $(\rep{1},\rep{1};\rep{20};d=2)$ under
$\alg{sl}(2,\Complex)$, $\alg{su}(4)$ and dilatations.
It is annihilated by $\gen{K},\gen{S},\gen{\bar S}$ 
and hence it is the primary field for a 
multiplet of local operators, cf.\ \figref{fig:Multi}.
This multiplet is very important because 
it contains all the conserved currents for $\superN=4$ SYM:
the $\alg{su}(4)$ Noether current $\Op{J}^m_{\mu n}$
transforming as $(\rep{2},\rep{2};\rep{15};d=3)$,
the supersymmetry currents $\Op{S}^a_{\mu\gamma},\Op{\bar S}_{\mu b\dot\gamma}$
transforming as $(\rep{3},\rep{2};\rep{4};d=3.5)$
and $(\rep{2},\rep{3};\rep{\bar 4};d=3.5)$
and the energy-momentum tensor $\Op{T}_{\mu\nu}$
transforming as $(\rep{3},\rep{3};\rep{1};d=4)$.
The currents define all Noether charges for $\alg{psu}(2,2|4)$, e.g.\
\[
R^a{}_b\sim\int d^3x\,\Op{J}^m_{0 n},\qquad
Q^a{}_{\gamma}\sim\int d^3x\,\Op{S}^a_{0 \gamma},\qquad
P_\mu\sim\int d^3x\,\Op{T}_{0 \mu}.
\]
Moreover, the multiplet contains two scalars
$\Op{L}\indup{kin},\Op{L}\indup{top}$  of dimension $d=4$. 
These are exactly the parity-even kinetic
and parity-odd topological parts of the Lagrangian density
\[
\Op{L}\indup{kin}=-\quarter\Gamma^{\mu\nu}\Gamma_{\mu\nu}+\half\partial^\mu\Phi_m\partial_\mu\Phi_m+\ldots,
\qquad
\Op{L}\indup{top}=\sfrac{1}{8}\varepsilon^{\mu\nu\rho\sigma}\Gamma_{\mu\nu}\Gamma_{\rho\sigma}.
\]
%

Next let us consider the labelling of representations for local operators. 
A lowest-weight representation is characterised by its primary field. 
The latter is characterised by the $\alg{sl}(2,\Complex)$ spin,
the $\alg{su}(4)$ representation and the conformal dimension $d$.
For instance the primary field $\Phi_m$ of the fundamental field representation
transforms as $(\rep{1},\rep{1};\rep{6};d=1)$ 
while the primary field $\Op{O}_{mn}$ of the energy-momentum representation 
transforms as $(\rep{1},\rep{1};\rep{20};d=2)$.
This characterisation is analogous 
to the discussion of unitary representations
of $\alg{su}(4)$ in \secref{sec:Algebra}.
The only difference is that the representation on 
the Taylor expansion of local operators is \emph{not unitary}:%
\footnote{Thanks to Gleb Arutyunov and Stefan Fredenhagen for helpful discussions regarding this issue. 
See also \cite{Janik:2010gc} for implications on
local operators, correlation functions, string states and their duality.} 
Surely $\alg{sl}(2,\Complex)$ has no finite-dimensional unitary
representations and also the dilatation generator $\gen{D}$ 
has \emph{imaginary eigenvalues}, cf.\ \eqref{eq:ConfTaylor}.
The point is that the Taylor components are not normalisable 
in the scalar product defining unitarity.
Nevertheless, there is a one-to-one map 
between representations for local operators and unitary representations.
It uses the following \emph{complex} conformal transformation of Minkowski space
\[
(t,x,y,z)\mapsto 2r^{-1}(iw,x,y,z)\mbox{ with }
w=1-\quarter x\mathord{\cdot} x\mbox{ and }
r=1-it+\quarter x\mathord{\cdot} x.
\]
It maps the dilatation generator
to $\gen{D}\mapsto i\gen{H}$
where $\gen{H}$ is the generator of the decompactified
$\alg{u}(1)$ discussed in \secref{sec:Algebra},
so the scaling dimension $d$ maps to the 
energy eigenvalue $E$.
Also the Lorentz algebra $\alg{sl}(2,\Complex)$ is
mapped to $\alg{su}(2)\oplus\alg{su}(2)$
which is commonly used to classify the
spin of fields in four dimensions.
For all practical purposes 
the complex nature of the above conformal transformation
is harmless in a perturbative quantum field theory
where one commonly continues into complex time directions anyway.
Therefore one often works with a dilatation generator
$\gen{D}'=-i\gen{D}$ whose spectrum is real 
and with $\alg{su}(2)\oplus\alg{su}(2)$ Lorentz generators
$\gen{L}'$ and $\gen{\bar L}'$.
Hence one can classify multiplets of local operators through
unitary representation of $\alg{psu}(2,2|4)$.
For instance the fundamental field 
and energy-momentum multiplets 
have Dynkin labels $[0;0;0,1,0;0;0]$ 
and $[0;0;0,2,0;0;0]$, respectively.
Note that, the representations $[0;0;0,p,0;0;0]$ are 
exceptionally short; the lowest state is annihilated by 
(at least) half of the supertranslations and hence the multiplet
is called \emph{half-BPS}.

\paragraph{Multiplet Splitting.}

Scaling dimensions $d$ for unitary representations 
can take arbitrary real values above a certain unitarity bound, 
cf.\ \eqref{eq:UniConstr}. 
Therefore, the scaling dimension 
typically varies smoothly with the coupling constant of the quantum theory.
However, representations at the lower bounds \eqref{eq:UniConstr}
have fewer components in general.
For example, the scaling dimension for half-BPS representations $[0;0;0,p,0;0;0]$
is fixed to $d=p$ and cannot depend on the coupling.

Nevertheless, there is an option to combine two or more 
short representations at the lower bound into a long representation
whose scaling dimension can then be increased smoothly.
This process called \emph{multiplet joining} 
(or \emph{multiplet splitting} in reverse)
is an analog of the Higgs effect where a massless vector
particle combines with a massless scalar particle to form
a massive vector. The set of local operators in $\superN=4$ SYM
has the exceptional feature that almost all short multiplets
of the classical theory can be combined into long multiplets
in the quantum theory. Only few short multiplets have no partner
(such as all half-BPS multiplets $[0;0;0,p,0;0;0]$)
and their scaling dimensions are therefore protected
from quantum corrections.

\begin{figure}\centering
\includegraphics{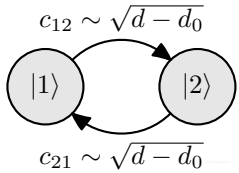}
\caption{Multiplet splitting at the unitarity bound.}
\label{fig:Splitting}
\end{figure}
Multiplet splitting takes place at the unitarity bound,
cf.\ \figref{fig:Splitting}:
Consider a long multiplet which decomposes into two short multiplets at $d=d_0$.
The representation of some generator $\gen{J}$ 
acts on states $\state{1},\state{2}$ of the submultiplets qualitatively as follows
\[
\gen{J}\state{1}=c_{12}\state{2}+\ldots,\qquad
\gen{J}\state{2}=c_{21}\state{1}+\ldots.
\]
The algebra relations imply that $c_{12}c_{21}\sim (d-d_0)$ because
splitting at $d=d_0$ requires $c_{12}=0$ or $c_{21}=0$.
Unitarity furthermore implies $c_{12}\sim c_{21}^\ast$
hence $c_{12}\sim c_{21}\sim \sqrt{d-d_0}$.
Therefore at $d=d_0$ the reality properties 
of the representation necessarily change,
i.e.\ $d\geq d_0$ is a unitarity bound.

\section{Isometries of the \texorpdfstring{$AdS_5\times S^5$}{AdS5 x S5} Superspace}
\label{sec:Isometries}

Supersymmetric strings require a ten-dimensional supergravity background 
as the space on which they can consistently propagate.
Next to a flat spacetime there exist two more maximal 
supersymmetric backgrounds. One of them is the $AdS_5\times S^5$ superspace.
According to the AdS/CFT correspondence
this string theory is exactly dual to conformal $\superN=4$ SYM on Minkowski space
being the boundary of $AdS_5\times S^5$, see \cite{Aharony:1999ti} for an extended review.
In the following we shall discuss this superspace, its boundary and its isometries
which are generated by the algebra $\alg{psu}(2,2|4)$.

\paragraph{\texorpdfstring{$AdS$}{AdS} Spacetime.}

We start by defining the anti de Sitter spacetime $AdS_{n+1}$ 
leaving $n$ generic for the time being. 
This $(n+1)$-dimensional spacetime has homogeneous negative curvature 
in close analogy to hyperbolic space $H^{n+1}$.
Similar space(time)s with homogeneous positive curvature
are the de Sitter spacetime $dS_{n+1}$ and the sphere $S^{n+1}$ (to which we shall frequently contrast $AdS_{n+1}$). 
There are several equivalent constructions which we shall now review.
One can embed it into $\Reals^{n,2}$ as single-shell hyperboloid  
specified by 
\[
AdS_{n+1}=\bigsetspec{X\in\Reals^{n,2}}{X\mathord{\cdot} X=-1},\qquad
S^{n+1}=\bigsetspec{Y\in\Reals^{n+2}}{Y\mathord{\cdot} Y=+1}.
\]
The metric is induced from the flat metric on $\Reals^{n,2}$
losing one time-like direction due to the condition $X\mathord{\cdot} X=-1$. 
An obvious alternative description uses time-like rays $[X]$ in $\Reals^{n,2}$
\[\label{eq:rays}
AdS_{n+1}=\bigsetspec{[X]}{X\in\Reals^{n,2},\,X\mathord{\cdot} X<0},
\quad\mbox{where }[X]=[Y] \mbox{ iff } X=zY\mbox{ with }z\in \Reals^+.
\]
The points $X$ or rays $[X]$ transform canonically under $\grp{SO}(n,2)$
and they are stabilised by a $\grp{SO}(n,1)$ subgroup. 
Consequently, $AdS_{n+1}$ can be viewed as the coset space
\[
AdS_{n+1}=\grp{SO}(n,2)/\grp{SO}(n,1),\qquad
S^{n+1}=\grp{SO}(n+2)/\grp{SO}(n+1).
\]
Thus the group of isometries of $AdS_5$ is $\grp{SO}(4,2)$.
Due to the presence of fermions, one should
promote the orthogonal to spin groups. 
For $n=4$ the group identities 
$\grp{Spin}(4,2)=\grp{SU}(2,2)$ and 
$\grp{Spin}(4,1)=\grp{Sp}(1,1)$  
furthermore allow to write
\[\label{eq:AdS5S5}
AdS_{5}=\grp{SU}(2,2)/\grp{Sp}(1,1),\qquad
S^{5}=\grp{SU}(4)/\grp{Sp}(2).
\]
%

\paragraph{Coordinates.}

There exist several choices of coordinates on $AdS_{n+1}$
which are useful in different situations. 
One is an analog of angle coordinates on the sphere:
Using trigonometric functions it is straight-forward to
construct a vector $X\in \Reals^{n,2}$ with $X\mathord{\cdot} X=-1$
(we shall use the signature $--+\ldots+$)
\[
\label{eq:AngleCoords}
X=(\sec\sigma\,\cos\tau,\sec\sigma\,\sin\tau,\tan\sigma\,\Omega),
\]
where $\Omega\in S^{n-1}\subset \Reals^{n}$ is a unit vector
and $\rho\in [0,\half\pi)$, $\tau\in[0,2\pi)$.
The induced metric reads
\[
\label{eq:AngleMetric}
ds^2=\sec^2\sigma\, (d\sigma^2-d\tau^2)+\tan^2\sigma\, d\Omega^2.
\]
On the coordinates $\Omega$ and $\tau$ the maximal compact subgroup
$\grp{SO}(n)\times\grp{SO}(2)$ acts canonically. 
The remaining $2n$ directions of $\grp{SO}(n,2)$ act non-trivially. 

A useful alternative is Poincar\'e-type coordinates $x\in \Reals^{n-1,1}$,
$y\in\Reals^+$ with the $\Reals^{n,2}$ embedding
\[
\label{eq:PoincareCoords}
X=y^{-1}\bigbrk{\half(x\mathord{\cdot} x+y^2+1),x,\half(x\mathord{\cdot} x+y^2-1)}.
\]
These coordinates reveal the conformally flat nature of the $AdS_{n+1}$ metric
\[
\label{eq:PoincareMetric}
ds^2=y^{-2}(dx\mathord{\cdot} dx+dy^2).
\]
A Poincar\'e subgroup of $\grp{SO}(n,2)$ acts 
on the $x$ while the corresponding dilatations
act as simultaneous scaling of $x$ and $y$
by the same factor. 
Special conformal transformations mix up $x$ and $y$ non-trivially
\[
\delta x\sim
x(\epsilon\mathord{\cdot} x)-
\half \epsilon(x\mathord{\cdot} x+y^2),\qquad
\delta y\sim
y(\epsilon\mathord{\cdot} x).
\]

Finally we note that isometries of $AdS_{n+1}$ also include reflections 
in $\Reals^{n,2}$.
For example, a reflection in the first component of the above $X$
corresponds to an inversion of time $\tau$ or
a conformal inversion of the coordinates $(x,y)\in \Reals^{n,1}$
\[
\tau\mapsto \pi-\tau,\qquad
(x,y)\mapsto -\frac{(x,y)}{x\mathord{\cdot} x+y^2}\,.
\]
%

\paragraph{Universal Cover.}

\begin{figure}\centering
\includegraphics[width=\textwidth]{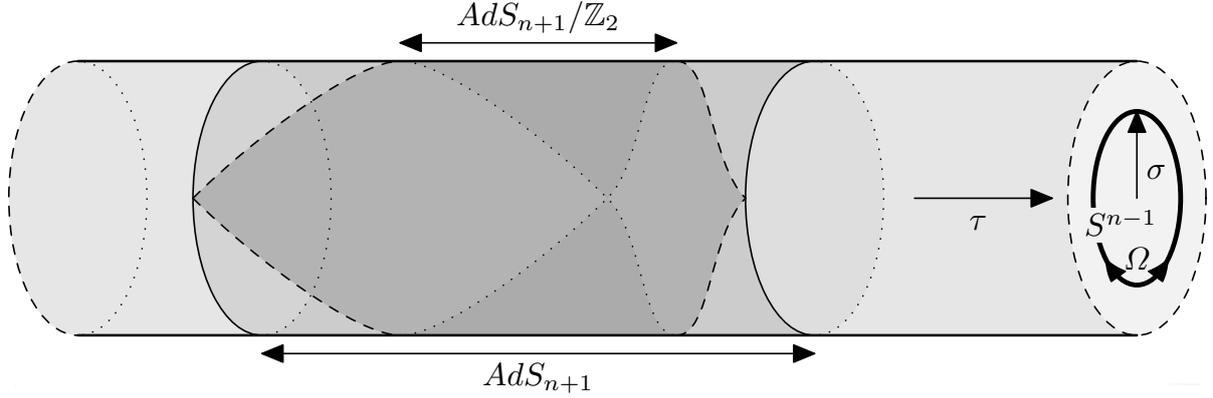}
\caption{Anti de Sitter space. 
The infinitely extended solid cylinder 
represents the universal cover $\widetilde{AdS}_{n+1}$
(light grey).
$AdS_{n+1}$ is obtained by identifying segments of 
time $\mathrm{\Delta}\tau=2\pi$ (medium grey).
The Poincar\'e patch $AdS_{n+1}/\Integers_2$ 
covers half of $AdS_{n+1}$ (dark grey).
The boundary $\partial \widetilde{AdS}_{n+1}=\Reals\times S^{n-1}$ 
is the outer shell of the cylinder.}
\label{fig:AdS}
\end{figure}

In \eqref{eq:AngleCoords} it is clear that the time coordinate $\tau$ 
is periodic: $\tau\equiv \tau+2\pi$.
Closed time-like curves are inconvenient for physical applications,
but luckily they can be removed by lifting 
to the \emph{universal cover} $\widetilde{AdS}_{n+1}$
on which a physical model can be defined. 
Hence the coordinates $(\tau,\sigma,\Omega)$ 
with non-periodic $\tau\in\Reals$ 
define a global chart for $\widetilde{AdS}_{n+1}$
which has the topology of an infinitely extended solid cylinder,
see \figref{fig:AdS}.
The natural embedding into $\Reals^{n,2}$ identifies
$\tau$ with $\tau+2\pi \Integers$ and leads to $AdS_{n+1}$.
Moreover, the Poincar\'e-type coordinates in \eqref{eq:PoincareCoords}
cover only half of $AdS_{n+1}$.
More precisely, 
if $\theta$ is the angle between $\Omega$ and $\Omega_0=(0,\ldots,0,1)$,
then the Poincar\'e patch is a wedge of the cylinder around $\tau=0$
defined by the inequality $\cos\tau>\sin\sigma \cos\theta$,
cf.\ \figref{fig:AdS}.

The universal cover $\widetilde{AdS}_{n+1}$ 
also has a direct formulation as a coset: 
The groups $\grp{SO}(n,2)$ and $\grp{SU}(2,2)$ have non-trivial coverings 
because their maximal compact subgroups contain 
the non-simply connected factors $\grp{SO}(2)$ and $\grp{U}(1)$, respectively.
The covering of $AdS_{n+1}$ is thus defined as 
\[
\widetilde{AdS}_{n+1}=\widetilde{\grp{SO}}(n,2)/\grp{Spin}(n,1),
\qquad
\widetilde{AdS}_{5}=\widetilde{\grp{SU}}(2,2)/\grp{Sp}(1,1).
\]
The universal covers $\widetilde{\grp{SO}}(n,2)$ and $\widetilde{\grp{SU}}(2,2)$
are physically relevant because they allow representations with
arbitrary real energies as compared to integer values
for $\grp{SO}(n,2)$ and $\grp{SU}(2,2)$.

\paragraph{Boundary of \texorpdfstring{$AdS$}{AdS}.}

The boundary $\partial AdS_{n+1}$ of $AdS_{n+1}$ is a $n$-dimensional spacetime.
It can be viewed as the space of light-like rays $[X]$ in $\Reals^{n,2}$,
cf.\ \eqref{eq:rays}
\[\label{eq:adsbound}
\partial AdS_{n+1}=\bigsetspec{[X]}{X\in\Reals^{n,2},\,X\mathord{\cdot} X=0},
\quad\mbox{where }[X]=[Y] \mbox{ iff } X=zY\mbox{ with }z\in \Reals^+.
\]
In the above coordinates of $AdS_{n+1}$ 
it is located at $\sigma=\half\pi$ or at $y=0$.
From \eqref{eq:adsbound} the topology of the $AdS_{n+1}$ boundary 
follows
\[
\partial AdS_{n+1}=S^{1}\times S^{n-1},\qquad
\partial \widetilde{AdS}_{n+1}=\Reals\times S^{n-1}.
\]
While the topology $S^{1}$ of time in $\partial AdS_{n+1}$ is periodic,
the boundary of the universal cover $\widetilde{AdS}_{n+1}$ 
has no closed time-like curves. 
Consequently it is the outer shell of the solid cylinder $\widetilde{AdS}_{n+1}$. 
The metric of $\Reals^{n,2}$ can be used to measure angles,
but not distances on the boundary, 
hence it merely induces a \emph{conformal} metric 
on $\partial AdS_{n+1}$
\[
ds^2\simeq -d\tau^2+d\Omega^2\simeq dx\mathord{\cdot} dx.
\]
In other words the boundary is conformally flat.
This is manifest in the Poincar\'e coordinates \eqref{eq:PoincareCoords} 
$x\in\Reals^{n-1,1}$ (with $y=0$)
on which $\widetilde{\grp{SO}}(n,2)$ acts 
by conformal transformations 
\eqref{eq:ConfCoord}.

Note that the boundary is at infinite distance 
to all points of $AdS_{n+1}$
(similarly to hyperbolic space $H^{n+1}$ and its
boundary $\partial H^{n+1}=S^{n}$).
Nevertheless the boundary can interact
with the bulk at finite times:
A light ray originating from $\sigma=\tau=0$ 
reaches the boundary $\sigma=\half\pi$ at $\tau=\half\pi$, 
cf.\ \eqref{eq:AngleMetric}. 
From there it travels back to the point $\sigma=0$ at time $\tau=\pi$.

\paragraph{\texorpdfstring{$AdS_5\times S^5$}{AdS5 x S5} Superspace.}

The $AdS_5\times S^5$ superspace is an extension of
$\widetilde{AdS}_5$ and $S^5$ by 32 fermionic directions.
It is very conveniently expressed as a coset space:
The groups $\grp{SU}(2,2)$ and $\grp{SU}(4)$ 
for the definition $AdS_5$ and $S^5$ in \eqref{eq:AdS5S5}
combine into the supergroup $\grp{PSU}(2,2|4)$ 
which has 32 fermionic directions.
Dividing by the bosonic denominator groups in \eqref{eq:AdS5S5}
one obtains the full superspace
\[
\widetilde{AdS}_5\times S^5\times \Complex^{0|16}
=\frac{\widetilde{\grp{PSU}}(2,2|4)}{\grp{Sp}(1,1)\times\grp{Sp}(2)}\,.
\]
The curvature radii of the $\widetilde{AdS}_5$ and $S^5$ subspaces
are equal but opposite, such that the overall
scalar curvature vanishes.

In view of the AdS/CFT correspondence, 
we shall consider the boundary of this superspace. 
The sphere $S^5$ is closed and the fermionic space $\Complex^{0|16}$ 
has trivial topology such that the overall boundary originates from
the $\widetilde{AdS}_5$ factor alone.
In the spherical coordinates \eqref{eq:AngleCoords},
it resides at $\sigma=\half\pi$. 
Let us approach the boundary 
with a codimension-one surface 
at a fixed $\sigma$ near $\sigma=\half\pi$. 
This surface has the topology $\Reals\times S^3\times S^5\times \Complex^{0|16}$.
According to \eqref{eq:AngleMetric}
the radius of the $S^3$ is $\tan\sigma$ 
while the radius of the $S^5$ factor is constantly $1$. 
Hence at the boundary the $S^5$ shrinks to a point
in comparison to the $S^3$.
This means that, for some physical purposes, the boundary of the $AdS_5\times S^5$
spacetime is effectively the boundary of $\widetilde{AdS}_5$ alone, 
i.e.\ $\Reals\times S^3$.
(A patch of) this spacetime is conformally equivalent to
Minkowski space $\Reals^{3,1}$.
The boundary of the $AdS_5\times S^5$ superspace 
has additional fermionic coordinates
to make up a conformally flat $\superN=4$ superspace.

\paragraph{Coset Space Sigma Model.}

In string theory isometries of the background spacetime 
become conserved Noether charges. 
This becomes obvious in the construction of a coset space sigma model, 
see the chapter \cite{chapSigma}.
Thus the group of global symmetries
of superstrings on $AdS_5\times S^5$ is
$\widetilde{\grp{PSU}}(2,2|4)$.
It should be noted that the 
coset space sigma model construction 
not only provides
the correct target space metric, 
but also a non-trivial superspace torsion 
and five-form supergravity flux 
coupling to the string worldsheet.

The $AdS_5\times S^5$ coset has a couple of exceptional 
features which make it a suitable background for a consistent quantum string theory:
First of all, it has 10 bosonic and 32 fermionic coordinates. 
Furthermore the worldsheet theory on this coset has 16 kappa symmetries
to reduce the effective number of fermionic coordinates to 16. 
Finally, the Killing form for $\grp{PSU}(2,2|4)$ vanishes
identically as required for conformal symmetry on the worldsheet. 
Only few cosets share these features, cf.\ \cite{Zarembo:2010sg}.

\phantomsection
\addcontentsline{toc}{section}{\refname}
\bibliography{chapters,conformal}
\bibliographystyle{nb}

\end{document}